\documentstyle[aps,preprint]{revtex}
\begin{document}
\draft

\preprint{KIAS-P98031, SNUTP98-122}

\title{BTZ black holes from the five-dimensional general
relativity with a negative cosmological constant}
\author{Youngjai Kiem$^{(a)}$\footnote{ykiem@kias.re.kr}
           and Dahl Park$^{(b)}$\footnote{dpark@ctp.snu.ac.kr} }
\address{ $^{(a)}$ School of Physics, KIAS, Seoul 130-012, Korea\\
          $^{(b)}$ Center for Theoretical Physics, Seoul National
                   University, Seoul 151-742, Korea}
\maketitle

\begin{abstract}
We show that the five-dimensional general relativity with 
a negative cosmological constant allows the solutions
of the form ${\cal M}_3 \times M_g$ where ${\cal M}_3$ is
the three-dimensional BTZ black hole and $M_g$ is a higher genus 
($g>1$) Riemann surface with a fixed size.  It is shown that
this type of spontaneous compactification on a Riemann surface
is possible only for the genus larger than one.  From type IIB
string theory point of view, certain near horizon geometry
of $D$ three-branes wrapped on the compact Riemann surface ($g>1$)
is the BTZ (or $AdS_3$) space-time tensored with the Riemann surface
and a constant size five-sphere.  The relevance of our analysis
to the positive energy conjecture of Horowitz and Myers is
discussed.
\end{abstract}

%\pacs{04.65.+e, 04.20.Jb, 04.50.+h}

\newpage

Ever since the realization that the Anti de Sitter (AdS) 
supergravity captures many aspects of the conformal field theory 
(CFT) on the boundary, there has been a growing interest toward 
the general relativity with a negative cosmological 
constant \cite{malda}.  
Of particular importance in this regard is the $AdS_5$/CFT$_4$ 
correspondence, in terms of which we can understand the 
strong coupling behavior of the four-dimensional supersymmetric
(or (2+1)-dimensional non-supersymmetric) Yang-Mills 
theory \cite{witten}.
Essentially asserting the stability of the Yang-Mills theory
via $AdS_5$/CFT$_4$ correspondence,
Horowitz and Myers proposed a physically pleasing but mathematically
vexing positive energy conjecture for the five-dimensional
general relativity with a negative cosmological constant;
it is conjectured that the energy functional of the smooth 
solutions of this theory that are asymptotically locally $AdS_5$
is bounded from below \cite{horowitz}.  By 
considering the general static solutions
of the form     
\begin{equation}
ds^2=g^{(2)}_{\alpha \beta} dx^{\alpha} dx^{\beta} 
+ e^{2\psi_1} d\theta^2 + e^{\psi_2} ( dx_1^2 + dx_2^2)
\label{torus}
\end{equation}
and obtaining the complete solution space for the space-time
geometry that is asymptotically $R \times S^1 \times T^2$, the
validity of the Horowitz-Myers conjecture in this limited 
case was demonstrated in Ref.~\cite{kp}.  The radial
and time coordinates span the two-dimensional subspace
with the metric $g^{(2)}_{\alpha \beta}$, $\theta$ represents
the circle $S^1$, and $(x_1 , x_2 )$ parameterizes the 
two-torus $T^2$.  Two exponential factors are the radial
coordinate dependent moduli fields for the circle size and
the two-torus size.  A natural generalization of the analysis
of \cite{kp} is to replace the two-torus with a Riemann 
surface of genus $g$\footnote{For the discussions on the
$M$ membranes wrapped on a higher genus Riemann surface, 
see \cite{emparan}.  See also \cite{mann}.}.  
Noting that there {\em exist} examples of
regular solutions with arbitrarily negative energy in the 
context of the asymptotically locally Minkowskian space-time 
in the genus zero case (bubble type solutions) \cite{brill}, 
the Horowitz-Myers conjecture suggests that the presence 
of the negative cosmological constant should drastically
influence the genus-dependent behavior of the solution 
space, so as to avoid the dangerous bubble type solutions
of the general relativity with zero cosmological constant.

In this paper, we find precisely such an example in the case
of the spontaneous compactification \cite{duff} 
of the five-dimensional
gravity to three dimensions upon the compactification
on a fixed size Riemann surface of an arbitrary genus
$g$.  Unlike the five-dimensional general relativity 
with zero cosmological constant where the three-dimensional
Minkowskian space-time tensored with a fixed-size two-torus
(genus one) is obvious allowed, we explicitly show that the 
type of the compactification in the above is possible only for the
Riemann surfaces with genus {\em larger} than one. 
In what follows, we consider the spontaneous compactification
on a fixed size Riemann surface with genus $g$ of the 
five-dimensional general relativity with a negative cosmological 
constant and get the general static radial dependent solutions.
For the genus larger than one, we show that the solutions
are three-dimensional Banados-Teitelboim-Zanelli (BTZ) black holes 
\cite{btz} tensored with the Riemann surface.  
For the genus zero and one 
case, on the other hand, this type of solutions do not exist.  
As will be explicitly demonstrated below, the solutions that we 
find from the five-dimensional analysis also become the solutions 
of the type IIB supergravity with a non-vanishing Ramond-Ramond (RR) 
self-dual five-form field strength, thereby related to 
$D$ three-branes.  The appearance of the BTZ black holes from
three-branes is an interesting feature resulting from the
compactification on a Riemann surface.
  
We start from the five-dimensional Einstein-Hilbert action with a 
negative cosmological constant
\begin{equation}
I= \int d^5x \sqrt{-g^{(5)}} \left( R^{(5)}+\frac{6}{l^2} \right) ,
\label{5action}
\end{equation}
where $R^{(5)}$ is the five-dimensional scalar curvature and $l$ is 
a positive constant.  This action can also be obtained from the 
type IIB supergravity upon the spontaneous compactification
on an $S^5$ in the presence of $D$ three-branes \cite{kp}.  As such,
the prototypical examples of the solutions of (\ref{5action}) are
$AdS_5$ space-time, $AdS_5$ black holes, and $AdS_5$ solitons,
all parallel to the world-volume of the $D$ three-branes.  We
instead consider a different topology, namely, ${\cal M}_3 \times
M_g$ where ${\cal M}_3$ is a three-dimensional space-time and 
$M_g$ is a genus $g$ Riemann surface.  For the genus zero surface,
we use the standard spherical metric $ds^2 = 
dx_1^2 + \sin^2 ( x_1 ) dx_2$, which produces a compact two-manifold.
For the genus one surface, the flat $R^2$ covering space metric
$ds^2 = dx_1^2 + dx_2^2$ can be used.  Upon taking the quotients
of the covering space by its discrete isometry subgroup
that acts freely, we get a compact two-torus.  For the higher 
genus case, we can use the hyperbolic metric 
$ds^2 = dx_1^2 + \sinh^2 (x_1 ) dx_2^2$; by 
taking the quotients of its universal covering space by its 
discrete isometry subgroup that acts freely but appropriately
discontinuously, we obtain a compact Riemann surface with any genus
larger than one.  Thus, the five dimensional metric is assumed
to be of the form
\begin{equation}
ds^2=g^{(2)}_{\alpha \beta} dx^{\alpha} dx^{\beta} + e^{2\psi_1} d\theta^2
+e^{\psi_2} ( dx_1^2+S^2(x_1) dx_2^2),
\label{5metric}
\end{equation}
similar to Eq.~(\ref{torus}) except that the two-torus part
has been replaced with a genus $g$ Riemann surface.
The function $S(x_1)$ is $\sin(x_1/k)$, 1 and $\sinh(x_1/k)$ for
the genus $g=0$, $g=1$ and $g>1$ case, respectively.  The constant
$k$ is so included that we can change the size of the Riemann
surface.  

Since we are interested in static, radial coordinate
dependent solutions, we assume that the two-dimensional metric 
$g^{(2)}_{\alpha \beta}$, the radius $\exp(\psi_1)$ of $S^1$ 
and the size $\exp(\psi_2)$ of $M_g$ depend only on the 
coordinates $x^{\alpha}$ (in fact, a space-like combination
of two $x^{\alpha}$s).  Under these assumptions, the non-trivial
part of the five-dimensional equations of motion from
the action (\ref{5action}) can be
equivalently summarized by
the following two-dimensional action:
\begin{equation}
I_2=\int d^2x \sqrt{-g} e^{-2\bar{\psi}} \left[ R+ \frac{6}{l^2}
e^{\bar{\psi}-\psi_1}-2 \mu e^{3\bar{\psi}}-\frac{3}{2}(D\psi_1)^2 
\right],
\label{2action}
\end{equation}
where $-2\bar{\psi}=\psi_1+\psi_2$ and the two-dimensional metric
has been rescaled as $g^{(2)}_{\alpha \beta}
=e^{-(3\psi_1+\psi_2)/2} g_{\alpha \beta}$.  The second term
in Eq.~(\ref{2action}) is the potential term originating from 
the five-dimensional cosmological constant, while the third
term comes from the curvature of the Riemann surface.  The
factor $\mu=-1/k^2$, $\mu=0$ and $\mu=1/k^2$ for the genus 
$g=0$, $g=1$ and $g>1$ case, respectively.  As expected, 
the $g=0$, $g=1$ and $g>1$ case contribute a negative, zero
and positive amount to the total potential energy, respectively,
$g=0$ case being the ``lowest energy $s$-wave sector''.  

We choose a radial gauge for the two-dimensional metric
\begin{equation}
ds^2=g_{\alpha \beta} dx^{\alpha} dx^{\beta}=
-\alpha^2 dt^2+\beta^2 dr^2 .
\label{2metric}
\end{equation}
One gets the static equations of motion from the two-dimensional action
Eq.~(\ref{2action}) under the choice of the radial gauge 
Eq.~(\ref{2metric}) by requiring that all the fields depend only 
on a space-like coordinate $r$. After a change of the space-like 
coordinate from $r$ to $x$ via
\begin{equation}
dx \equiv \frac{\sqrt{2}}{l} \alpha \beta e^{-\psi_1-\bar{\psi}} dr,
\label{defx}
\end{equation}
the static equations of motion are
summarized by a one-dimensional action
\begin{equation}
I_1= \int dx \left[ \alpha \Omega^{1/2} e^{-\psi_1} \alpha' \Omega'
-\frac{3}{4} \alpha^2 \Omega^{3/2} e^{-\psi_1}  \psi_1^{\prime 2}
-\frac{\mu l^2}{2} \Omega^{-1} e^{\psi_1} \right] , 
\label{1action}
\end{equation}
where $\Omega=e^{-2\bar{\psi}}$ and the prime denotes the differentiation
with respect to $x$. The gauge constraint
\begin{equation}
\alpha \Omega^{1/2} e^{-\psi_1} \alpha' \Omega'
-\frac{3}{4} \alpha^2 \Omega^{3/2} e^{-\psi_1}  \psi_1^{\prime 2}
+\frac{\mu l^2}{2} \Omega^{-1} e^{\psi_1}=\frac{3}{2}
\label{gconst}
\end{equation}
should be supplemented to the equations of motion from the 
action (\ref{1action}).  We note that the field $\beta$ is 
completely absorbed into the space-like coordinate $x$, 
Eq.~(\ref{defx}), so that $\beta$ does not appear
in the one-dimensional action (\ref{1action}) and the 
gauge constraint (\ref{gconst}).  This is possible since
the field $\beta$ in Eq.~(\ref{2metric}), that depends 
only on $r$, can be completely gauged away by 
a diffeomorphism of $r$.  We have to determine three
functions $\Omega$, $\psi_1$ and $\alpha$ by solving the
equations of motion from (\ref{1action}).  For this purpose,
we observe that there are three symmetries of the action 
(\ref{1action}):
\begin{eqnarray*}
&(a)&~ x \rightarrow x + \epsilon \\
&(b)&~ \psi_1 \rightarrow \psi_1+\epsilon, 
 ~x \rightarrow e^{-\epsilon} x \\
 &(c)&~ x \rightarrow e^{\epsilon} x, 
 ~\Omega \rightarrow e^{\epsilon} \Omega,
 ~\alpha \rightarrow e^{-\epsilon/4} \alpha ,
\end{eqnarray*}
where $\epsilon$ is an arbitrary real parameter of each symmetry 
transformation. Using these three 
symmetries, we can integrate the coupled second order differential
equations once to get coupled first order differential equations 
simply by constructing three Noether charges
\begin{eqnarray}
c_0&=&\alpha \Omega^{1/2} e^{-\psi_1} \alpha' \Omega'
-\frac{3}{4} \alpha^2 \Omega^{3/2} e^{-\psi_1}  \psi_1^{\prime 2}
+\frac{\mu l^2}{2} \Omega^{-1} e^{\psi_1} 
\label{c0} \\
\psi_0-c_0 x &=& -\frac{3}{2} \alpha^2 \Omega^{3/2} e^{-\psi_1}
\psi_1'
\label{psi0} \\
s+c_0 x &=& \alpha \Omega^{3/2} e^{-\psi_1} \alpha'
-\frac{1}{4} \alpha^2 \Omega^{1/2} e^{-\psi_1} \Omega' ,
\label{s}
\end{eqnarray}
corresponding to each symmetry. The gauge constraint Eq. (\ref{gconst})
dictates that $c_0=3/2$, and the total number of constants of motion 
thus reduces from six to five.

We are interested in the case where the size $\exp (\psi_2 )$ of 
$M_g$ is fixed 
\begin{equation}
e^{\psi_2}=\Omega e^{-\psi_1}=e^{\psi_{20}} ,
\label{psi1}
\end{equation}
where $\psi_{20}$ is a constant.  This condition determines 
the solution for $\psi_1$ in terms of $\Omega$ and one 
constant of motion $\psi_{20}$ and, as such, the total number 
of constants of motion reduces from five to four.
Plugging Eq.~(\ref{psi1}) and Eq.~(\ref{psi0}) into Eq.~(\ref{c0}),
we get
\[
\frac{d}{dx} \ln | \alpha^2 \Omega^{-3/2} | =
 \frac{3-\mu l^2 e^{-\psi_{20}}}{x-2 \psi_0 /3} ,
\]
which can be integrated to yield
\begin{equation}
\alpha^2=\alpha_0^2 | x-2 \psi_0 /3 |^{3-\mu l^2 \exp(-\psi_{20})}
\Omega^{3/2} ,
\label{alpha}
\end{equation}
where $\alpha_0$ is a constant of integration.
Using Eq.~(\ref{psi1}) and Eq.~(\ref{alpha}), we can rewrite 
Eq.~(\ref{psi0})
\[
\alpha_0^2 e^{\psi_{20}} \Omega \Omega'
= (x-2 \psi_0 /3) | x-2 \psi_0 /3|^{\mu l^2 \exp(-\psi_{20})-3},
\]
which, upon integration, becomes
\begin{equation}
\alpha_0^2 e^{\psi_{20}} \Omega^2=\frac{2}{\mu l^2 e^{-\psi_{20}}-1}
| x-2 \psi_0 /3|^{\mu l^2 \exp(-\psi_{20})-1}+c_1 ,
\label{Omega}
\end{equation}
where $c_1$ is a constant of integration.  When $\mu l^2 e^{-\psi_{20}}=1$
such that the denominator of the right hand side of Eq.~(\ref{Omega}) 
vanishes, the solution (\ref{Omega}) changes to a logarithmic function.  

We determined all the fields without using Eq.~(\ref{s}) and, therefore,
the consistent solutions for the spontaneous compactification
on a fixed size Riemann surface should satisfy Eq.~(\ref{s}). 
Plugging the solutions Eqs.~(\ref{psi1}), (\ref{alpha}) and
(\ref{Omega}) into Eq.~(\ref{s}) yields a consistency condition
\begin{equation}
s+\frac{3}{2}x=\frac{5-\mu l^2 e^{-\psi_{20}}}{2 \mu l^2 e^{-\psi_{20}}-2}
\left(x-\frac{2}{3}\psi_0\right)+\frac{\epsilon_x c_1}{2}
 (3-\mu l^2 e^{-\psi_{20}})
\left|x-\frac{2}{3}\psi_0\right|^{2-\mu l^2 \exp(-\psi_{20})} ,
\label{g2const}
\end{equation}
where $\epsilon_x$ is the sign of $(x-2 \psi_0 /3)$.
Since Eq.~(\ref{g2const}) should be satisfied for all values of 
$x$, the power of the second term on the right hand side should be 
either zero or one, or the second term itself should vanish
by requiring $c_1 (3-\mu l^2 \exp (-\psi_{20} ))= 0$.  The power of 
one is inconsistent; when
the power is one, $\mu l^2 e^{-\psi_{20}}=1$, the first term
on the right hand side becomes a logarithmic function 
proportional to $\ln |x-2 \psi_0 /3|$, while the left hand
side remains unchanged (see the remarks below Eq.~(\ref{Omega})).
When the power is zero
or equivalently when 
\begin{equation}
\mu l^2 e^{-\psi_{20}}=2 ,
\label{mu}
\end{equation}
Eq.~(\ref{g2const}) reduces to
\begin{equation}
c_1=2\epsilon_x (s+\psi_0),
\label{c1}
\end{equation}
determining the constant of motion $c_1$ in terms of others.
When $c_1 = 0$, matching the coefficient of $x$ in Eq.~(\ref{g2const})
yields the condition (\ref{mu}).  In this case, matching the constant 
terms of Eq.~(\ref{g2const}) gives $ s + \psi_0 = 0$, which yields 
$c_1 = 0$ again according to Eq.~(\ref{c1}).
When $\mu l^2 \exp (-\psi_{20} ) = 3$, the coefficient of 
$x$ in Eq.~(\ref{g2const}) can not match, and thus this
case is inconsistent.  Therefore, we
conclude that Eq.~(\ref{mu}) is the consistency condition 
and $c_1$ is determined by Eq.~(\ref{c1}).
As argued before, as a result, there are four constants 
of motion parameterizing the solution space, $\psi_{20}$,
$\alpha_0$, $\psi_0$ and $s$.  The key point is that 
Eq.~(\ref{mu}) requires that $\mu$ should be positive definite.
The genus $g=0$ and $g=1$ cases thus can not satisfy 
the equations of motion, namely Eq.~(\ref{s}).  In other words,
the spontaneous compactification on a fixed size Riemann surface
is only possible for the compact manifold whose universal 
covering space metric is the hyperbolic type (thereby producing 
a negative two-dimensional intrinsic curvature) or the Riemann
surface with genus higher than one.  This illustrates one of 
our main points that
the solution space of the five-dimensional general relativity
with a negative cosmological constant has a drastic dependence
on the genus.  More concretely, for the compactification on
a {\em fixed size} flat torus, one can explicitly compute 
the five-dimensional Ricci tensor and find 
$R_{x_1 x_1} = R_{x_2 x_2 } = 0$, contradicting the 
five-dimensional field equations.  Similarly,
for the genus zero Riemann surface, even if the Ricci tensor 
components transversal to the Riemann surface, 
$R_{x_1 x_1}$ and $R_{x_2 x_2}$, are proportional to the 
metric tensor, the sign of the proportionality constant is
opposite to the sign of the proportionality constant for
the components longitudinal to the Riemann surface, resulting
another contradiction to the five-dimensional field equations. 
At least in the case considered in this paper, there are no
consistent solutions for the would-be ``lower''
energy sector of genus zero.
 
To summarize, the consistent solutions for the higher genus 
Riemann surface compactification are 
\begin{eqnarray}
e^{\psi_1}&=&e^{-\psi_{20}}\Omega
\label{psi1f} \\
\alpha^2&=&\alpha_0^2 | x-2 \psi_0 /3 | \Omega^{3/2}
\label{alphaf} \\
\alpha_0^2 e^{\psi_{20}} \Omega^2&=&2|x-2\psi_0/3|+2\epsilon_x (s+\psi_0)
\label{Omegaf} \\
\mu \equiv \frac{1}{k^2} &=&\frac{2}{l^2}e^{\psi_{20}}~~~(g>1)
\label{muf}
\end{eqnarray}
with four constants of motion, $\psi_{20}$, $\alpha_0$, $\psi_0$ 
and $s$.  The solutions Eqs.~(\ref{psi1f}) - (\ref{muf}) in fact
correspond to the three-dimensional BTZ black holes (the zero mass
black hole, positive mass black holes and the $M=-1$ solution)
tensored with a constant size Riemann surface, 
as we will see shortly.  Even if we 
have two sets of solutions depending on the value of 
$\epsilon_x=\pm 1$, it turns out that the two sets of 
solutions describe the same space-time.  To avoid redundancy, 
we hereafter restrict our presentation to the case of $\epsilon_x=+1$.
The possible range of $x$ is determined from Eq.~(\ref{alphaf}) 
and Eq.~(\ref{Omegaf}) as
\begin{equation}
x-2\psi_0 /3 >0,~~~x-2\psi_0 /3+(s+\psi_0) >0.
\label{xrange}
\end{equation}
For the more palpable comparison to the analysis of Horowitz and
Myers on the higher than three dimensional gravity theories with
a negative cosmological constant \cite{horowitz}, we consider 
the two cases when $s+\psi_0>0$ and when $s+ \psi_0 < 0$.  
First, when $s + \psi_0 >0$, the second condition of 
Eq.~(\ref{xrange}) is
automatically satisfied if the first condition of Eq.~(\ref{xrange})
holds. A convenient choice of a radial coordinate $r$ is
\[
\frac{r^2}{l^2}- r_0^2 =x- \frac{2}{3}\psi_0 >0
\]
where $r_0^2 = s +\psi_0$.
Using this radial coordinate, we can write the five-dimensional 
metric as
\begin{equation}
ds^2= -\left(\frac{r^2}{l^2}- r_0^2 \right) dt^2 
+\frac{r^2}{l^2} d \theta^2 + \left(\frac{r^2}{l^2}- r_0^2 \right)^{-1} 
dr^2 +dx_1^2 + \sinh^2 \left( \frac{\sqrt{2}x_1}{l} \right)
dx_2^2 , 
\label{metric1}
\end{equation}
where we fixed two constants of integration as $e^{\psi_{20}}
=1$ and $\alpha_0^2=2$, which correspond to the scale choice 
for the circle and the $M_g$.  Of the original four constants of 
integration, these two scale choices fix two of them. Since 
$\psi_0$ can be deleted from the metric by an appropriate 
diffeomorphism, we disregard it as a gauge dependent parameter. 
Thus we end up with one positive semi-definite constant, $r_0^2$, 
that parameterize the solution space in a gauge-independent 
fashion.  We also rescaled the time coordinate as 
$2dt^2 \rightarrow dt^2$.  The metric (\ref{metric1}) is the
familiar metric for the BTZ black holes tensored
with a constant size Riemann surface \cite{btz}.  Secondly, when 
$s+\psi_0<0$, the first condition of Eq.~(\ref{xrange}) is
automatically satisfied if the second condition of Eq.~(\ref{xrange})
holds.  A convenient choice of a radial coordinate $r$ is
\[
\frac{r^2}{l^2} - r_0^2 =x-2\psi_0 /3+(s+\psi_0) >0
\]
where $r_0^2 =-(s+\psi_0)$.  Using this radial coordinate, 
we can write the five-dimensional metric as
\begin{equation}
ds^2= - \frac{r^2}{l^2} dt^2+ \left(\frac{r^2}{l^2}-r_0^2 \right) 
d \theta^2  +  \left(\frac{r^2}{l^2} - r_0^2 
\right)^{-1} dr^2 +dx_1^2 + \sinh^2 \left( \frac{\sqrt{2}x_1}{l} \right)
dx_2^2
\label{metric2}
\end{equation}
where we also imposed the conditions $e^{\psi_{20}}=1$ and
$\alpha_0^2=2$, fixing two constants of integration as before,
and $ 2dt^2 \rightarrow dt^2$.  
The metric (\ref{metric2}) is now seen to be the double Wick rotated 
version of the metric (\ref{metric1}), just as what happens in the
higher dimensional analysis of the Horowitz and Myers \cite{horowitz}.  
To avoid the conical singularity at $r/l = r_0$, the 
value of $r_0$ for the
metric (\ref{metric2}) should be set to a particular value
depending on the asymptotic size of $\theta$ coordinate.
Then, the space-time is everywhere regular and we can interpret 
it as the $AdS_3$ solitons tensored with a Riemann 
surface \cite{horowitz}.  
Based on the fermion holonomy in the
boundary at infinity, it was identified in \cite{horowitz} that
the asymptotically (locally) $AdS_5$ analog of the metric (\ref{metric1}) 
belongs to the Ramond sector (R sector)
and the asymptotically (locally) $AdS_5$ analog of
the metric (\ref{metric2}) belongs to the lower energy 
Neveu-Schwarz sector (NS sector) from the higher dimensional
analysis.  In fact, via a coordinate transformation
$r^2 / l^2 \rightarrow r^2 / l^2 + r_0^2$, the metric 
(\ref{metric2}) can be rewritten as
\begin{equation}
ds^2= -\left(\frac{r^2}{l^2}+ r_0^2 \right) dt^2 
+\frac{r^2}{l^2} d \theta^2 + \left(\frac{r^2}{l^2}
+ r_0^2 \right)^{-1} dr^2
 +dx_1^2 + \sinh^2 \left( \frac{\sqrt{2}x_1}{l} \right)
dx_2^2 .
\label{metric3}
\end{equation}
This corresponds to the negative mass ($M= -1$) BTZ solution
tensored with a Riemann surface \cite{btz}.  From the purely
three-dimensional bulk analysis it has also been considered 
as the NS sector due to the antiperiodicity of the Killing 
spinors \cite{carlip}, consistent with the assignment 
of \cite{horowitz}.  Due to the existence of the Killing spinors
for three-dimensional part of the solutions Eq.~(\ref{metric3}) 
(and thus the solutions Eq.~(\ref{metric2})) \cite{carlip}, the 
three-dimensional $AdS_3$ solitons are guaranteed to have 
the stable minimum energy, its energy being lower than that of
the R sector vacuum with periodic Killing spinors, the massless 
BTZ black holes \cite{henn}.
The main difference between the 
three-dimensional and a higher dimensional
case is that the interconnecting solutions of \cite{russo} and
\cite{kp}  with naked 
singularities that interconnect the R and NS sectors do not exist for 
the three dimensional (non-rotating) case.  The structure of the
bosonic regular solution space is similar for both cases, with
a finite mass gap between the R sector and the NS sector.  

We note that the two sets of solutions, Eq.~(\ref{metric1}) and 
Eq.~(\ref{metric2}), are also the solutions of the ten-dimensional 
type IIB supergravity.  The bosonic type IIB supergravity 
action appropriate for our context is 
\begin{equation}
I_{10}= \int d^{10}x \sqrt{-g^{(10)}}\left[ e^{-2\phi}
(R^{(10)}+4(D\phi)^2)-\frac{1}{2 \cdot 5!}F^{(10)2} \right] ,
\label{10action}
\end{equation}
where $\phi$ is the dilaton field, $F^{(10)}$ is the RR self-dual 
five-form field strength and $g_{A B}^{(10)}$, the 
ten-dimensional metric.  The self-duality condition
for the five-form field strength should be imposed by 
hand \cite{townsend}.  The equations of motion from 
the action (\ref{10action}) are
\begin{equation}
e^{2\phi} D^{\mu}( 4e^{-2\phi} D_{\mu} \phi )+ R^{(10)}+4(D\phi)^2=0
\label{dilaten}
\end{equation}
for the dilaton field,
\begin{equation}
D^M F^{(10)}_{MABCD}=0
\label{rr}
\end{equation}
for the RR gauge field and 
\begin{equation}
R^{(10)AB}=e^{2\phi} T^{AB}
\label{gg}
\end{equation}
for the ten-dimensional metric, where the stress-energy tensor $T^{AB}$ 
is obtained by varying
\[
L=4 e^{-2\phi} (D\phi)^2 -\frac{1}{2 \cdot 5!} F^{(10)2}
\]
with respect to the ten-dimensional metric.
The ten-dimensional self-dual solutions from our five-dimensional
analysis can be written as
\begin{eqnarray}
ds^2&=&g^{(10)}_{AB} dx^A dx^B=g^{(5)}_{\mu \nu}dx^{\mu}dx^{\nu}
+ (\sqrt{2} l)^2 d\Omega^2_5  ,
\label{10metric} \\
\phi&=& {\rm constant} , 
\label{dilaton} \\
F^{(10)}_{\mu_1 \mu_2 \mu_3 \mu_4 \mu_5}&=& 32 l^{-1} e^{-2 \phi} 
\epsilon_{\mu_1 \mu_2 \mu_3 \mu_4 \mu_5} ,
\nonumber \\
F^{(10)}_{\theta_1 \theta_2 \theta_3 \theta_4 \theta_5}&=&
 32 l^{-1} e^{-2 \phi} (\sqrt{2} l)^5 
\epsilon_{\theta_1 \theta_2 \theta_3 \theta_4 \theta_5} ,
\label{5form}
\end{eqnarray}
where $g^{(5)}_{\mu \nu}$ is the five dimensional metric,
Eq. (\ref{metric1}) or Eq. (\ref{metric2}), and the metric 
$d\Omega_5^2$ is the 
standard metric on the unit $S^5$ with five angle coordinates
$\theta_1$, $\theta_2$, $\theta_3$, $\theta_4$ and $\theta_5$.
The antisymmetric tensor $\epsilon_{\mu_1 \mu_2 \mu_3 \mu_4 \mu_5}$ 
is the volume form on the five-manifold and 
$\epsilon_{\theta_1 \theta_2 \theta_3 \theta_4 \theta_5}$ is the
volume form on the unit five-sphere.  The solutions 
Eqs.~(\ref{10metric}) - (\ref{5form}) that satisfy
\[
R^{(10)}=R^{(5)}+\frac{10}{l^2}=0 ~~~, ~~~ D^M F^{(10)}_{MABCD}=0
\]
\[
R^{(10)}_{\mu \nu}=-\frac{2}{l^2}g^{(5)}_{\mu \nu}
~~,~~R^{(10)}_{\mu \theta_i}=0
~~,~~R^{(10)}_{\theta_i \theta_j}
=\frac{2}{l^2} g^{(10)}_{\theta_i \theta_j}~~~(i,j=1,\cdots,5)
\]
are actually the solutions of the ten-dimensional field equations
Eqs.~(\ref{dilaten}) - (\ref{gg}).  From the type IIB supergravity
point of view, our solution thus represents a certain near-horizon
geometry produced by the $D$-three branes wrapped on a compact 
Riemann surface with the genus $g>1$, and the space-time geometry 
is of the form ${\cal M}_3 \times M_g \times S^5$. Here, 
${\cal M}_3$ is the BTZ black hole whose $R$ sector ground
state is the massless black hole and the $NS$ sector ground state, 
the $AdS_3$ solitons. 
The wrapped three-branes becomes a string-like object and we end 
up getting the ground state geometry of the BTZ type.  An intriguing
point that deserves further understanding from the microscopic 
$D$ three-brane side is that this type of configuration is
possible only for the Riemann surfaces with the genus larger 
than one.   

From the analysis in this paper on the spontaneous compactification
of the five-dimensional general relativity with a negative
cosmological constant on a Riemann surface, we accumulated
some supporting evidences toward the validity of the
Horowitz-Myers conjecture \cite{horowitz}; our results suggests 
a possibility that the presence 
of the negative cosmological constant may prevent the 
existence of the bubble type solutions (genus zero solutions)
\cite{brill} that plagued the 
stability of the asymptotically locally Minkowskian gravity.
Furthermore, it was pointed out that the $M=-1$ solution of
the BTZ model has the higher dimensional analog that was 
independently argued to belong to the NS sector in \cite{horowitz}.
In particular, the R sector solutions and the $NS$ sector 
solutions of the BTZ model are related to each other 
by the double Wick rotation.    
It was suggested by Strominger that the three-dimensional 
BTZ black holes may capture the essentials of the higher 
dimensional stringy (non-extremal) black holes such as the 
five-dimensional $D1$-$D5$-$KK$ black holes of the type IIB 
supergravity compactified on a $T^5$ \cite{strominger}.  
A crucial element
in this type of argument is the existence of the $U$-dual 
transformation that takes the five-dimensional $D1$-$D5$-$KK$
black holes into the BTZ black hole \cite{hyun}.  Our results 
indicates that, if such a $U$-dual transformation is explicitly 
found in our context, the BTZ black holes may also be used to better 
understand the non-extremal black holes produced by
$D$ three-branes and vice versa.  Along with the microscopic 
explanation for the restriction of the Riemann surfaces to the 
genus larger than one, this issue in our opinion deserves 
further investigations.  
 
\acknowledgements{D.P. was supported by the Korea
Science and Engineering Foundation (KOSEF) through the Center for
Theoretical Physics (CTP) at Seoul National University (SNU).}

\end{document}